\documentclass[twocolumn,prl,a4paper,floatfix,preprintnumbers,amsmath,amssymb]{revtex4-1}
\usepackage{amssymb,graphicx}
\usepackage[utf8]{inputenc}

\newcommand{\mus}{\mu{\rm s}}
\newcommand{\kHz}{{\rm kHz}}

\usepackage{color}

\begin{document}

\title{Experimental realization of quantum zeno dynamics}

\author{
F.\ Sch\"{a}fer$^{1,2,\ast}$,
I.\ Herrera$^1$,
S.\ Cherukattil$^1$,
C.\ Lovecchio$^1$,
F.\ S.\ Cataliotti$^{1,2,3}$,\\
F.\ Caruso$^{1,2,3}$,
A.\ Smerzi$^{1,3,4}$
}

\affiliation{
$^1$LENS - Universit\`{a} di Firenze, Via Nello Carrara 1, 50019 Sesto Fiorentino, Italy\\
$^2$Dipartimento di Fisica ed Astronomia, Via Sansone 1, 50019 Sesto Fiorentino, Italy\\
$^3$QSTAR, Largo Enrico Fermi 2, 50125 Firenze, Italy\\
$^4$Istituto Nazionale di Ottica, INO-CNR, Largo Enrico Fermi 2, 50125 Firenze, Italy\\
$^\ast$schaefer@lens.unifi.it
}

\begin{abstract}
 It is generally impossible to probe a quantum system without
 disturbing it. However, it is possible to exploit the back-action
 of quantum measurements and strong couplings to tailor and protect
 the coherent evolution of a quantum system. This is a profound and
 counterintuitive phenomenon known as quantum Zeno dynamics (QZD).
 Here we demonstrate QZD with a rubidium Bose-Einstein condensate in
 a five-level Hilbert space. We harness measurements and strong
 couplings to dynamically disconnect different groups of quantum
 states and constrain the atoms to coherently evolve inside a two-level
 subregion. In parallel to the foundational importance due to the
 realization of a dynamical superselection rule and the theory of
 quantum measurements, this is an important step forward in
 protecting and controlling quantum dynamics and, broadly speaking,
 quantum information processing.
\end{abstract}

\maketitle

Back-action is a concept that lies at the very heart of quantum
mechanics. In contrast to classical mechanics, it is impossible to
probe a quantum system without disturbing it (unless the system is
in an eigenstate of the observable). Yet, quantum mechanics allows
the exploitation of back-actions in order to drive the dynamics
along different quantum paths. As first noticed by von
Neumann~\cite{neumann_mathematische_1932}, a given quantum state can
indeed be guided into any other state by tailoring a specific
sequence of measurements. When the measurements lead back to the
initial state and are frequent enough~\cite{smerzi_zeno_2012}, the
back-action freezes out the dynamics: a prominent example of
measurement-induced disturbance which later took the name of the
quantum Zeno effect (QZE)~\cite{misra_zenos_1977}. The QZE was first
experimentally proven in a closed system to significantly slow down
Rabi-driven oscillations between two
levels~\cite{itano_quantum_1990}. In open systems, the Zeno effect
as well as an acceleration due to an anti-Zeno
effect~\cite{kofman_2000, kofman_2001} were first demonstrated in
Ref.~\cite{fischer_2001}. Notice that both effects can also
influence thermalization~\cite{erez_2008}. A general approach for
the control of quantum behavior by coupling to the continuum was
presented in ref.~\cite{kofman_2001}. However, quantum Zeno does not
necessarily imply a freeze-out. Quite to the contrary, back-actions
can be also exploited to {\it preserve} a coherent dynamics in a
subspace of the Hilbert space and to forbid transitions among
engineered boundaries. This phenomenon is known as quantum Zeno
dynamics (QZD) and has been predicted in Ref.~\cite{facchi_2002}. A
possible implementation of QZD with atoms in optical cavities has
been theoretically investigated in Ref.~\cite{raimond_quantum_2012}.

Here, we experimentally realize QZD in a five-level Hilbert space
describing a Bose-Einstein condensate (BEC) of $^{87}$Rb atoms in a
magnetic micro-trap. In particular, by means of engineered strong
couplings and controlled measurements, different regions of the
Hilbert space can be dynamically disconnected, i.e.\ the transfer of
a physical state from one of this region to another is dynamically
forbidden. This corresponds to projecting the full Hamiltonian into
an effective one, with the latter involving only part of the Hilbert
space. Furthermore, the distinctive controllability of our physical
system allows us to apply four different coupling protocols and to
show how to confine $^{87}$Rb atoms in a two-level subregion, where
they evolve coherently. This is also tested via Ramsey
interferometry and successfully compared with a simple theoretical
model. Finally, we demonstrate that the four different coupling
protocols lead to the same QZD evolution. This constitutes also the
first experimental verification of the theoretical
prediction~\cite{facchi_quantum_2008} that QZD can be equivalently
attained (asymptotically) by frequent projective measurements,
strong continuous coupling, and fast unitary kicks.

\begin{figure}[t]
    \centering
   	    \includegraphics[width=80mm]{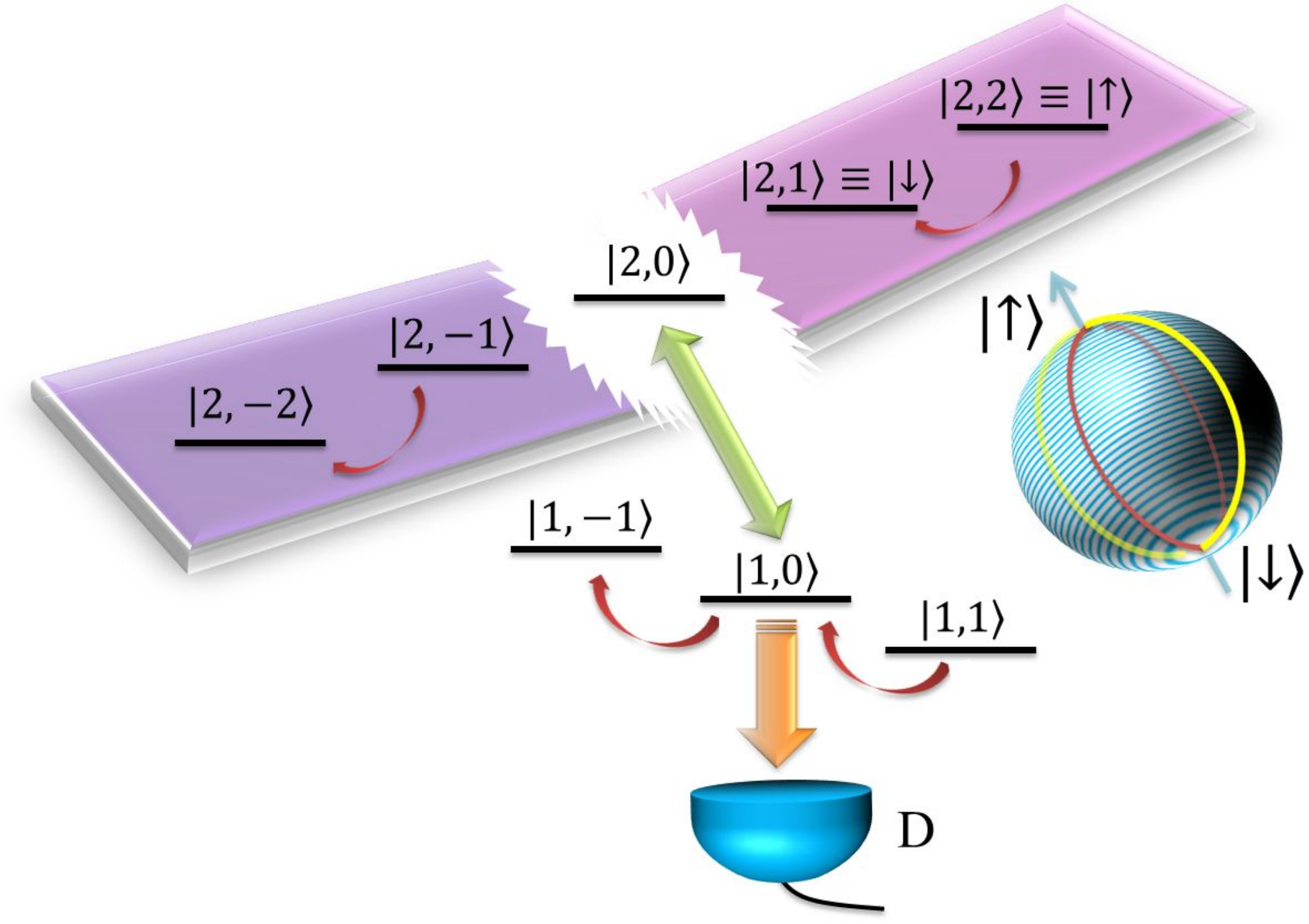}
		\caption{
			{\bf Representation of our QZD scheme.}
			Shown is the level structure of the $^{87}$Rb atom cloud
			that is initially in the $|F,m_F\rangle = |2,2\rangle$
			state. The levels of the protected two-level subspace are
			$|2,2\rangle$ and $|2,1\rangle$, denoted as
			$|\!\uparrow\rangle$ and $|\!\downarrow\rangle$,
			respectively. An applied RF field at $2.171~{\rm MHz}$
			couples neighbouring $m_F$ states (thin, red arrows), a
			laser induced Raman transition couples the $|2,0\rangle$ and
			$|1,0\rangle$ states (thick, green arrow). The hyperfine
			splitting between the $F=2$ and $F=1$ manifolds is about
			$6.834~{\rm GHz}$. An additional laser field connects the
			$F=1$ states to an external level (thick, orange arrow), and
			induces dissipation by spontaneous decay (lifetime $6~{\rm
			ns}$). This corresponds to the detection D. Rabi
			oscillations (around an $x$-axis) and Ramsey rotations
			($y$-axis) of the protected two-level subspace, discussed
			later in the text, are indicated on the Bloch
			sphere~\cite{nc}.
    }
    \label{fig:levels}
\end{figure}

\section{Results}

{\bf Experimental setup.}
In this Article we present experimental results obtained on a
Rubidium ($^{87}$Rb) BEC evolving in a five-level Hilbert space
given by the five spin orientations of the $F=2$ hyperfine ground
state (see Fig.~\ref{fig:levels}). We use a time sequence of
pulses to perform a state-selective negative measurement, namely, a
measurement of the \textit{absence} of population from one
particular level of the system. This dynamically decouples the
Hilbert space in two-level subregions. The atoms are found to
oscillate coherently between the states $|F, m_F \rangle =
|2,2\rangle \equiv |\!\uparrow\rangle$ and $|2,1\rangle \equiv
|\!\downarrow\rangle$ without leaking out to the other available
states. The same results have been obtained in absence of
measurements by exploiting unitary couplings, both constant and
pulsed in time. This proves the
prediction~\cite{facchi_quantum_2008} that measurements are not a
crucial ingredient for attaining QZD. We further prove via Ramsey
interferometry that the strong back-actions preserve, rather than
destroy, the coherence of the atomic two-level superposition.

The initial state of all the experiments presented in the following
is a cloud of $^{87}$Rb atoms in the $|\!\uparrow\rangle$ state Bose
condensed in a magnetic trap. After creation of the condensate (see
Methods for more details on the experimental procedure) the atoms
are released from the trap. A homogeneous magnetic bias field
defines the quantization axis of the system and lifts the degeneracy
of the states. We then shine a radio frequency (RF) field resonantly
coupling neighbouring $m_F$ states in the $F=2$ manifold.

\begin{figure}[t]
    \centering
			\includegraphics[width=80mm]{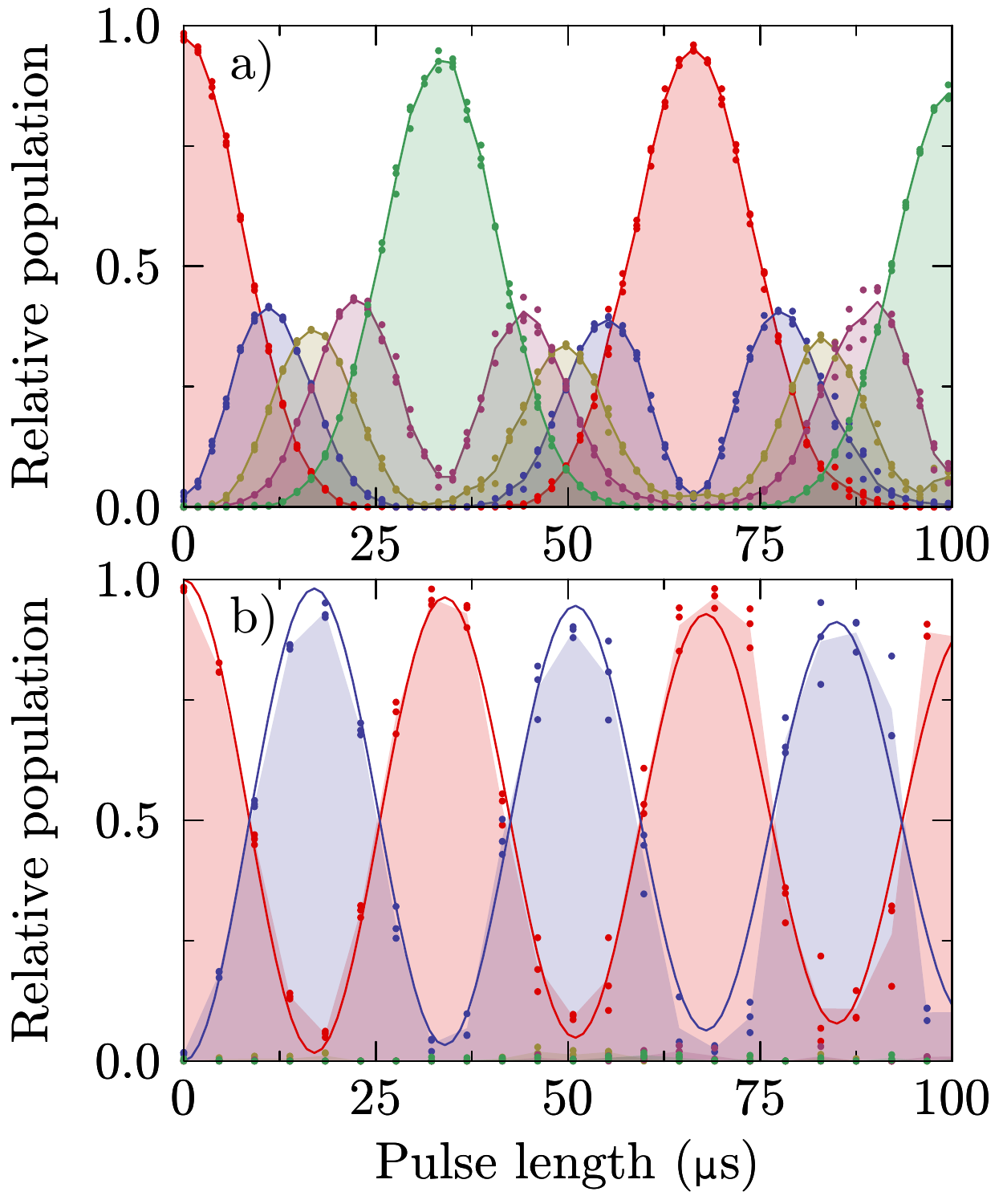}
    \caption{
			{\bf Rabi oscillations observed in the $F=2$ manifold.}
			(a) With only the RF field applied. Experimental data are
			represented by dots. As there is not sufficient data for
			statistically significant error bars we instead report all
			the data from three runs of the experiment. The shaded areas
			correspond to their average values, while the lines are the
			model predictions of Eq.~(\ref{eqn:H}). The relative
			populations ($m_F = +2$: red; $+1$: blue; $0$: yellow; $-1$:
			violet; $-2$: green) are reported as function of the RF pulse
			length. The observed oscillations involve all five states. (b)
			With RF field plus constant Raman beams of coupling strength
			$2\pi\,200~\kHz$ applied. The Figure is organized as in (a)
			but for the solid lines that here represent the solution of
			Eq.~(\ref{eqn:density}). The populations oscillate between the
			$|2,2\rangle \equiv |\!\uparrow\rangle$ (red) and $|2,1\rangle
			\equiv |\!\downarrow\rangle$ (blue) states. The observed
			oscillations decay, because at finite Raman powers the atoms
			eventually leak out of the protected subspace and into the
			(here unobserved) $F = 1$ hyperfine states.}
    \label{fig:rabi_and_constRaman}
\end{figure}

Under the effect of the RF field the system evolves according to the
Hamiltonian
\begin{equation}
    H_{F=2} = \left(
    \begin{array}{ccccc}
        0 & \Omega & 0 & 0 & 0 \\
        \Omega & 0 & \sqrt{3/2} \ \Omega & 0 & 0 \\
        0 & \sqrt{3/2} \ \Omega & 0 & \sqrt{3/2} \ \Omega & 0 \\
        0 & 0 & \sqrt{3/2} \ \Omega & 0 & \Omega \\
        0 & 0 & 0 & \Omega & 0
    \end{array}
    \right)\, ,
    \label{eqn:H}
\end{equation}
where the basis is chosen to start from $m_F = +2$. It is obtained
by applying angular-momentum algebra and the rotating-wave
approximation~\cite{ketterle_evaporative_1996}. Here and in the
following we set $\hbar = 1$ and express all energies in angular
frequencies. The Rabi frequency $\Omega$ is proportional to the RF
field intensity. We choose $\Omega = 2\pi\,15~\kHz$. As a result of
this interaction, the atomic population cycles periodically between
all the five sub-levels of the $F=2$ manifold as shown in
Fig.~\ref{fig:rabi_and_constRaman}(a).

{\bf State selective measurements.}
We implement a state selective measurement as a two-step process: a
full population transfer from $|2,0\rangle$ to $|1,0\rangle$ is
followed by a population measurement of the latter. Technically
speaking, first two laser beams (``Raman beams'') induce a Raman
transition between $|2,0\rangle$ and $|1,0\rangle$. We then
illuminate the BEC with a laser beam (``Dissipative Light'')
resonant with the $F=1 \rightarrow F'=2$ transition to the $5
P_{3/2}$ excited state from which atoms will decay outside the BEC.
During the decay a photon is emitted~\cite{itano_quantum_1990}. This
process is commonly considered a measurement even if the emitted
photon is not detected. Note that this radiation will only affect
atoms in the $F=1$ state, i.e.\ atoms that were originally in $|2,
0\rangle$.

In order to show which ingredients are really mandatory to achieve
QZD we implement four different experimental protocols: In the first
one, we realize a series of discrete measurements by a periodic
application of our state selective measurement scheme. In the second
protocol, we apply continuously both the Raman beams and the
Dissipative Light and we perform a continuous measurement. In the
third and fourth protocols, no Dissipative Light is used and the
dynamics is unitary. In the third experiment unitary kicks are
realized by a sequence of $\pi$-pulses of the Raman beams. In the
fourth protocol, we implement a continuous coupling scheme and the
Raman beams are kept on continuously. See Methods for details on the
parameters chosen in each case.

In all these experiments, by means of the Raman beams we are
effectively observing the \emph{absence} of any population in the
$|2,0\rangle$ state. In the language of
QZD~\cite{facchi_quantum_2008}, this accounts to projecting our
Hamiltonian Eq.~(\ref{eqn:H}) into an effective one, $H'=P\,H\,P$,
\begin{eqnarray}
        H' &=& \left(
    \begin{array}{ccccc}
        0 & \Omega & 0 & 0 & 0 \\
        \Omega & 0 & 0 & 0 & 0 \\
        0 & 0 & 0 & 0 & 0 \\
        0 & 0 & 0 & 0 & \Omega \\
        0 & 0 & 0 & \Omega & 0
    \end{array}
    \right)\, \ \textrm{and} \   \\
		P &=& \left(
    \begin{array}{ccccc}
        1 & 0 & 0 & 0 & 0 \\
        0 & 1 & 0 & 0 & 0 \\
        0 & 0 & 0 & 0 & 0 \\
        0 & 0 & 0 & 1 & 0 \\
        0 & 0 & 0 & 0 & 1 \\
    \end{array}
    \right), \nonumber
\label{eqn:Hprime}
\end{eqnarray}
where $P$ is the projector on the subspace spanned by all sublevels
except the observed one, $|2,0\rangle$. As we start with all the
population in state $|\!\uparrow\rangle$, we expect the evolution to
be restricted to the two states $|\!\uparrow\rangle$ and
$|\!\downarrow\rangle$. A typical result obtained with our fourth
protocol (continuous unitary coupling) is reported in
Fig.~\ref{fig:rabi_and_constRaman}(b). In contrast to the case of
purely RF induced oscillations,
Fig.~\ref{fig:rabi_and_constRaman}(a), the population now oscillates
between the first two sub-levels only (rotations around $x$-axis in
the Bloch sphere~\cite{nc} in Fig.~\ref{fig:levels}) with nearly
negligible decay toward the other levels. This is the hallmark of
QZD.

\begin{figure}[t]
    \centering
    \includegraphics[width=80mm]{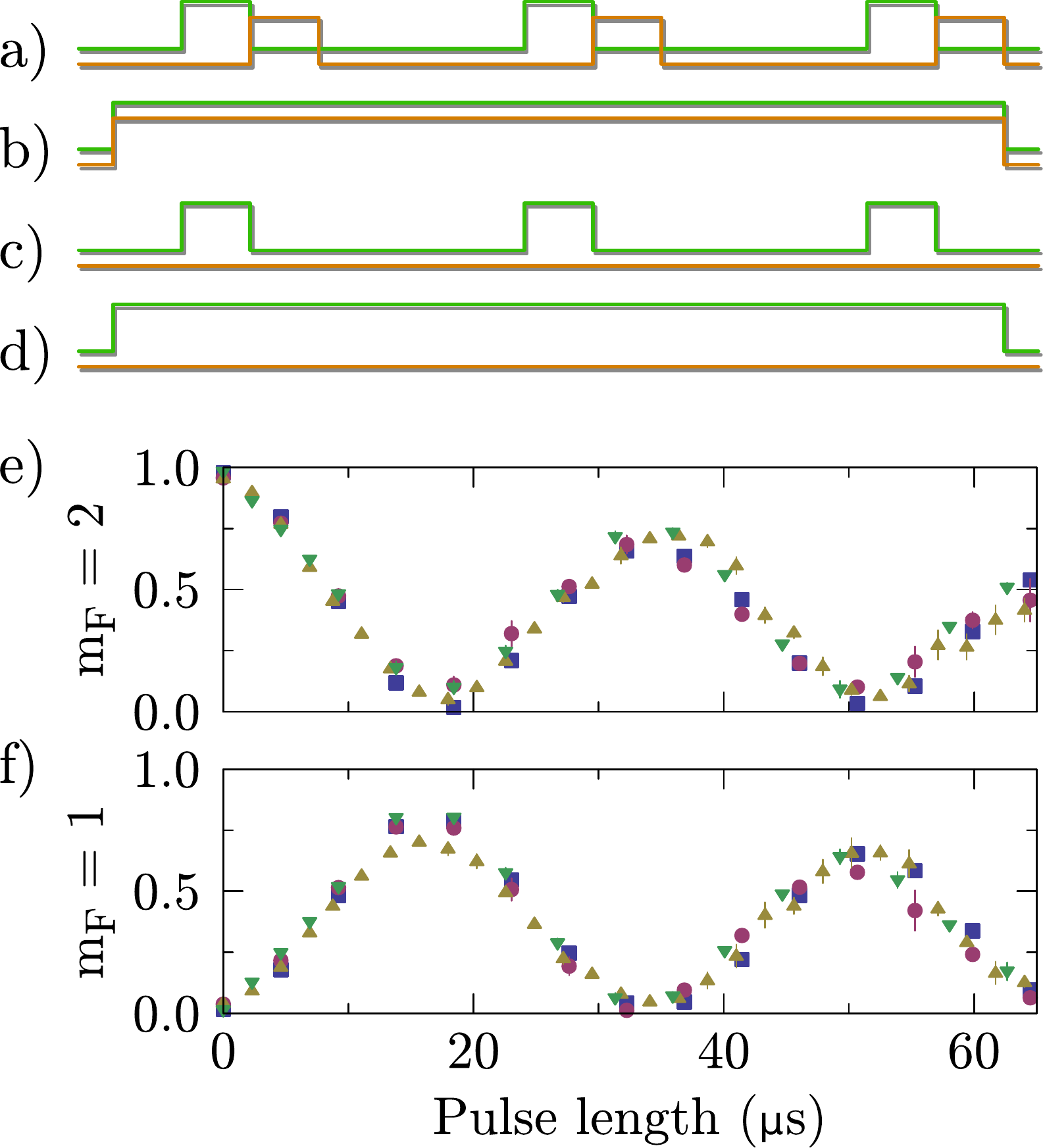}
    \caption{
			{\bf Comparison of four different protocols to induce QZD.}
			Shown is a comparison between RF induced Rabi oscillations for
			all four experimental protocols: a) discrete measurements
			(green down triangles), b) continuous measurements (violet
			circles), c) unitary kicks (yellow up triangles), and d)
			continuous unitary coupling (blue squares). See Methods
			section for details on the chosen parameters. Panel e) and f)
			show the relative populations in the $m_F=2$ and $m_F=1$
			states, respectively, as a function of total RF pulse length.
			We report the average populations with error bars (mostly
			comparable to the data symbol size) indicating the spread of
			the data. The lower four sets of lines illustrate the time
			sequences of the four protocols. The Raman beams are in green,
			the Dissipative Light is in orange. The horizontal time axis
			is not to scale.
		}
    \label{fig:comparison}
\end{figure}

{\bf Experimental results.}
A comparison of the results obtained with the four different
protocols is given in Fig.~\ref{fig:comparison}. For the chosen
parameters, we observe a similar population dynamics. This is in
close agreement with the prediction~\cite{facchi_quantum_2008} that
frequent projective measurements, strong continuous coupling or fast
unitary kicks should all asymptotically lead to a confined dynamics
according to Eq.~(\ref{eqn:Hprime}). As indicated by our
experimental results, the degree of confinement can be strong and
comparable (within the experimental error bars) even with finite
measurement rates and coupling strengths. This freedom is important
with respect to future applications of QZD.

In order to quantify the coherence of the dynamically created
two-level system we describe the observed oscillations obtained with
our fourth protocol with an effective two-state model (see
Methods). We introduce phenomenologically the rate
$\Gamma_{\rm loss}$ to account for losses from state
$|\!\downarrow\rangle$ to external states and $\gamma_{\rm deph}$ to
account for the loss of coherence between the states
$|\!\uparrow\rangle$ and $|\!\downarrow\rangle$. These two terms
have different physical origin. The rate $\Gamma_{\rm loss}$ arises
from the imperfect protection of the Zeno subspace due to our finite
coupling strength~\cite{streed_continuous_2006}.  The dephasing rate
$\gamma_{\rm deph}$ is originated in our experiment by the presence
of high frequency noise on the magnetic bias field and the RF
signal, and by off-resonant scattering of photons from the Raman
beams. This model is valid only for short evolution times. Due to
the finite size of our Hilbert space and the finite coupling
strength, the population can come back to the initial state,
$|\!\uparrow\rangle$. This means that for long times we experience a
revival of population into the Zeno (partially) protected subspace.
Therefore we apply our two-level model only for times up to
$100~\mus$, where the population revival is still not affecting the
observed dynamics.

\begin{figure}[t]
    \centering
			\includegraphics[width=80mm]{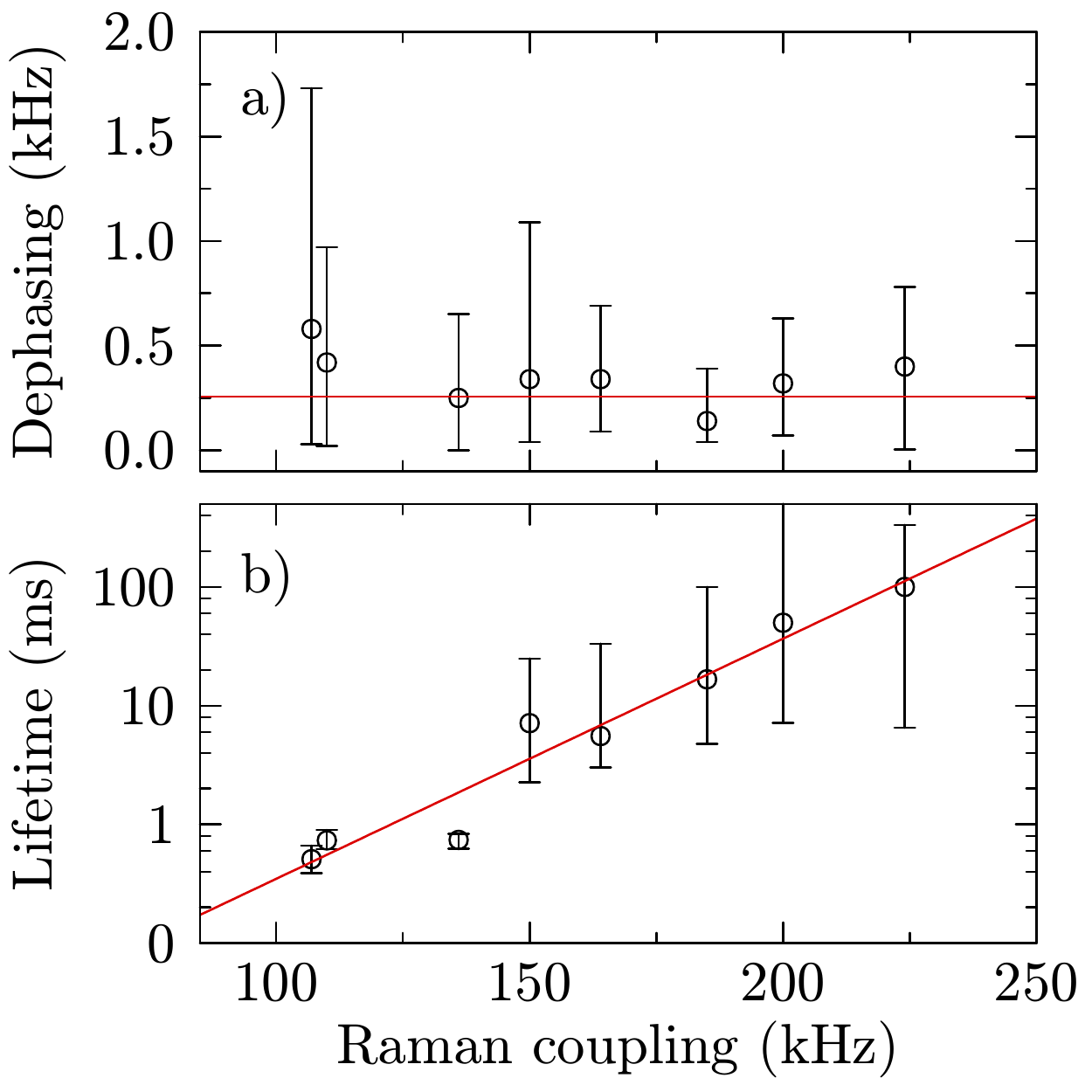}
    \caption{
			{\bf Dephasing and lifetimes in a protected subspace.}
			The dephasing rates (a) and lifetimes (b) of the atom
			population in the protected subspace under constant Raman
			beams illumination are reported. The factor $2\pi$ in the
			coupling frequencies is suppressed. The RF induced Rabi
			coupling is constant at $2\pi\,15~\kHz$. The data (points) are
			well fitted (red lines) by a constant dephasing rate and by an
			exponentially increasing lifetime, respectively. The error
			bars indicate the uncertainty in the parameter estimation (see
			Methods for details). The dephasing rates display only a weak
			dependence on the Raman coupling strength. The lifetimes are
			observed to increase hundredfold.
      }
    \label{fig:lifetime}
\end{figure}

In Fig.~\ref{fig:rabi_and_constRaman}(b) we included a fit of our
model to the experimental data of the population dynamics of the
states $|\!\uparrow\rangle$ and $|\!\downarrow\rangle$. From the fit
we extract the values $\gamma_{\rm deph} =
2\pi\,0.3_{-0.2}^{+0.3}~\kHz$ and $\Gamma_{\rm loss} = 2\pi\,
0.02_{-0.02}^{+0.12}~\kHz$, corresponding to a lifetime of the
particles in the subspace of $50~{\rm ms}$. In the upper panel of
Fig.~\ref{fig:lifetime} we report the dephasing times,
$2\pi/\gamma_{\rm deph}$, obtained by a scan of the Raman couplings
in the range from $2\pi\,100~\kHz$ to $2\pi\,225~\kHz$. Only a weak
dependence of the dephasing rate on the Raman coupling is observed.

A complete suppression of a leakage of atoms out of the two-level
subspace is only expected at infinite Raman coupling strength where
the projector $P$ of Eq.~(\ref{eqn:Hprime}) would be exactly
realized. In Fig.~\ref{fig:lifetime} (lower panel) we report the
lifetimes $2\pi/\Gamma_{\rm loss}$ obtained for different settings
of the Raman coupling strength. We observe a crossover from a full
five-level dynamics at $100~\kHz$ to a nearly pure two-level
dynamics above $200~\kHz$. In this range the lifetime increases
hundredfold at an exponential rate of about $0.05~\kHz^{-1}$.

To further investigate the coherence of the dynamics in the
two-level subspace we realized a Ramsey interferometric scheme
within the two-level subspace (corresponding to rotations around the
$y$-axis in the Bloch sphere in Fig.~\ref{fig:levels}). In the
presence of constant Raman beams two RF pulses are sent onto the
atoms separated by a variable delay time $T$. Each RF pulse
encompasses a $\pi/2$ area such that the first pulse evenly
populates the two states of our subspace. Note that this is
impossible in absence of the Raman beams as was already observed
in~\cite{minardi_time-domain_2001}. The populations after the second
pulse are then recorded as a function of $T$. An oscillatory
population distribution between the two sub-levels is expected if
and only if there is coherence between both
states~\cite{ramsey_experiments_1990}.

\begin{figure}[t]
    \centering
        \includegraphics[width=80mm]{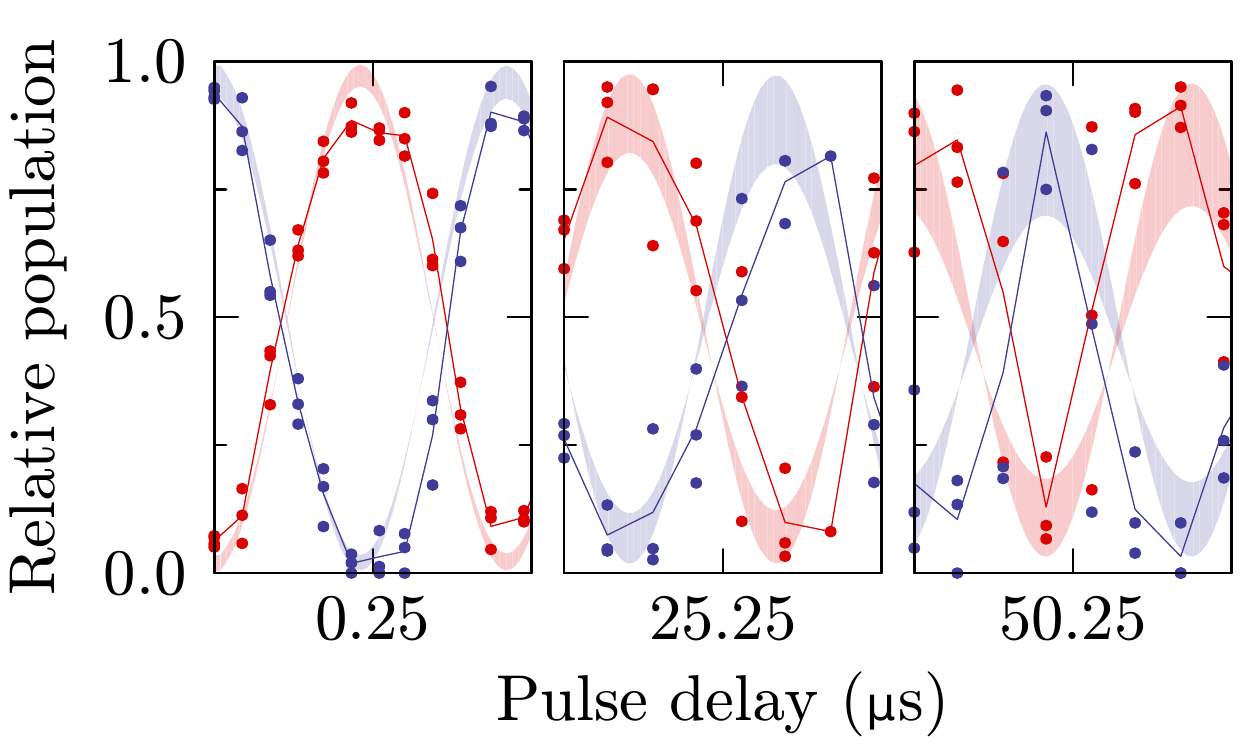}
    \caption{
			{\bf Demonstration of coherent dynamics.}
			A Ramsey scheme experiment reveals interference fringes as
			function of the delay time $T$ between the two RF pulses. For
			better resolution we show the data obtained only in three
			ranges of $T$. Each window has a temporal width of $0.5~\mus$.
			Reported are the populations of the $|\!\downarrow\rangle$
			(red) and $|\!\uparrow\rangle$ (blue) states. Shown are the
			raw data (points, three measurements per delay time),
			their mean values (connected straight lines) and the model
			prediction (shaded areas) with the corresponding parameters
			taken from Fig.~\ref{fig:lifetime}. The Raman beams induced
			a coupling strength of $2\pi\,140~\kHz$. In the observed
			pulse delay range only a small loss of fringe contrast is
			noted.
		}
    \label{fig:ramsey}
\end{figure}

We report our experimental findings in Fig.~\ref{fig:ramsey}. The
populations show clear oscillations with fringe contrast close to
unity over the whole observed pulse delay range $T$. Only a small
damping of the oscillations is perceivable. This concludingly
confirms that QZD preserves the coherence of the dynamics in
protected subspaces.

\section{Discussion}

We have demonstrated that measurements and strong couplings can be
tailored to create disjoint Hilbert subspaces. We have shown that a
two-level system initially localized in one of this subspaces will
evolve coherently without leaking probability to neighbouring
Hilbert subspaces. This realizes a dynamical superselection rule
which, in analogy to the ``W3" superselection rule for
charge~\cite{wigner_1952}, imposes that if two initial states are
separated in different regions of the Hilbert space, the separation
will persist at all times.

A general crucial problem of quantum information is to protect
experimental protocols from decoherence and probability leakage. A
strategy discussed in the literature is to encode the information on
a part of the overall Hilbert space where noise cancels out. The
existence of such decoherence-free subspaces was demonstrated in
Refs.~\cite{white2000,blatt}. Also, strategies for bath-optimized
protection of quantum information were discussed~\cite{clausen2010}.
Here, we provide the proof-of-principle of an alternative
noise-protection scheme which is realized via four different
protocols. The idea is to dynamically exploit QZD to drive the
dynamics inside a designed region of the full space of accessible
states and preserve quantum coherence against any leakage to the
environment. Notice that, in a different context and perspective
though, a broad variety of pulse control techniques have been
developed and used in NMR quantum computation to suppress
decoherence and confine the dynamics in
subsystems~\cite{viola_dynamical_1998,nmr} and, very recently, also
for nitrogen-vacancy centers in diamond~\cite{london2013}. A future
exploration of QZD in our five-level system will be the realization
of a controllable two-qubit gate, where strong couplings are
exploited to protect the two different qubits and tailor their
interaction. In view of possible applications one could implement
all sources necessary for our schemes in a single integrated device
by replacing the Raman beams with microwave radiation.

The possibility of engineering the coupling of the system to the
environment for quantum state preparation and quantum computation is
of particular relevance as shown theoretically in
Refs.~\cite{plenio1999,Cirac2009,Zoller2013,Knight2000}. Our
approach can be explored in other physical systems, such as quantum
cavities or trapped ions, realizing multi-level subspaces (qudits)
or continuous-variable ones. Therefore, our results open up the way
to the control of quantum information systems in the broader sense
of Quantum Zeno Dynamics.


\section{Methods}

\begin{figure}[t]
    \begin{center}
        \includegraphics[width=80mm]{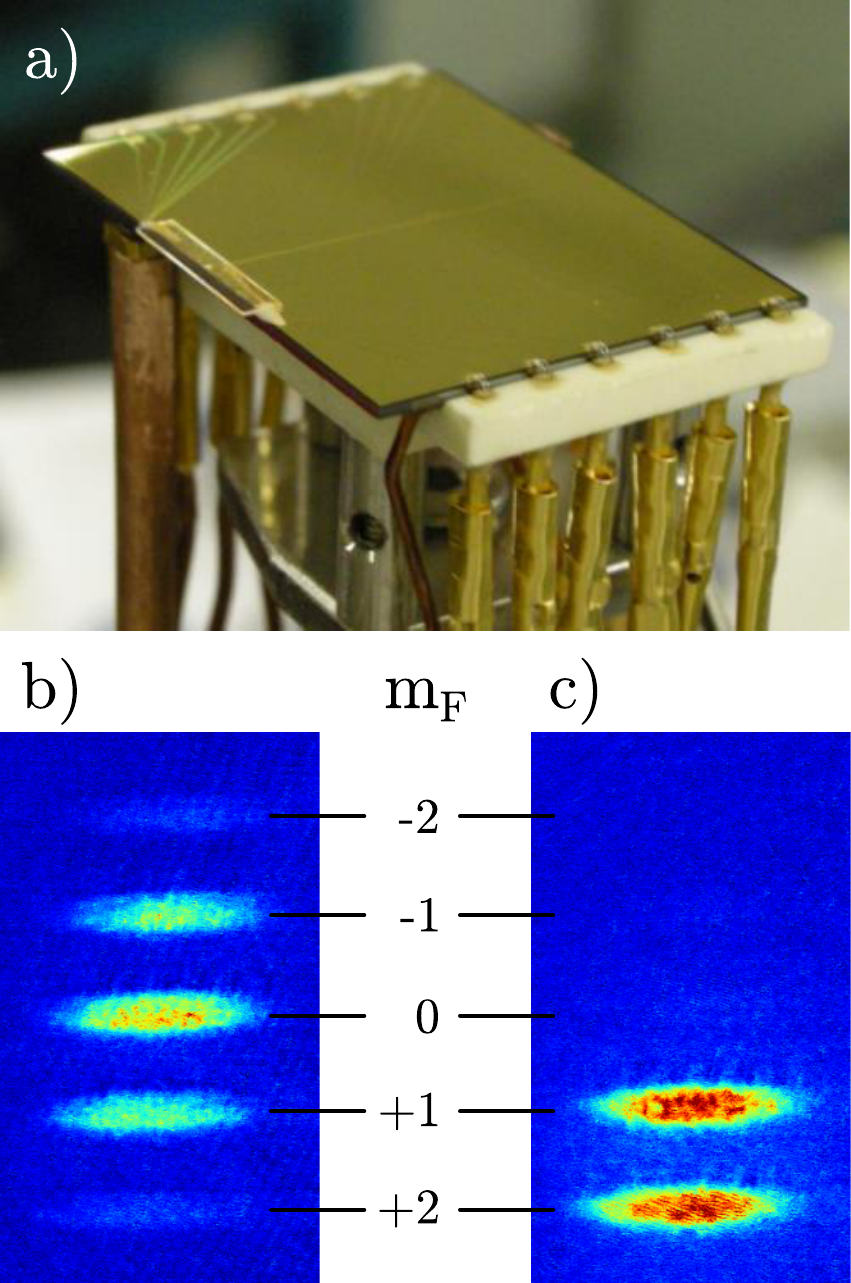}
    \end{center}
    \caption{
			{\bf Rb BEC on an atom chip.}
			a) Picture of the atom chip before it was installed inside the
			vacuum system. Visible is the chip ($23~\rm{mm} \times 30~{\rm
			mm}$) with its reflective, golden surface. On its left an
			embedded photonic structure (not used in the present
			experiment) is visible. Two rows of wires interface the chip
			structures with the experiment control. In the final setup the
			chip is mounted upside down. Panels b) and c) show absorption
			images taken $24~{\rm ms}$ after release from the magnetic
			trap. In b) is shown the population distribution after an RF
			$\pi/2$-pulse in the absence of the Raman beams. The
			population is centered around $|2,0\rangle$. Panel c) shows
			the situation with constant Raman beams. An appropriate RF
			$\pi/2$-pulse now even populates the $|\!\uparrow\rangle$ and
			$|\!\downarrow\rangle$ states. This configuration is used to
			initialize the Ramsey interferometer in the protected
			subspace.
			}
    \label{fig:si1}
\end{figure}

{\bf Creation of the $^{87}$Rb Bose-Einstein condensate.} We produce
a Bose-Einstein condensate (BEC) of $^{87}$Rb atoms in a magnetic
micro-trap realized with an atom chip (see upper part of
Fig.~\ref{fig:si1}). The apparatus consists of a single cell
only, for both, initial atom trapping and BEC creation. The main
component of the magnetic trap consists of a Z-shaped wire on the
atom chip with just $125~\mu{\rm m}$ diameter that hosts a current
of $1.8~{\rm A}$. The magnetic trap is completed by an external
pair of Helmholtz coils. The loading procedure starts by emission
of hot Rb atoms into the vacuum cell by means of a pulsed rubidium
dispenser. The atoms are then laser cooled in a magneto-optical
trap and transferred to the chip trap using an ancillary magnetic
trap realized by sending a current of 25 A through a wire ($1~{\rm
mm}$ diameter) placed behind the chip (not visible in
Fig.~\ref{fig:si1}). BEC is reached in $8~{\rm s}$ by forced
evaporative cooling, ramping down the frequency of a radio
frequency (RF) field. The RF fields for evaporation and
manipulation of the Zeeman states are produced by two further
conductors also integrated on the atom chip. The BEC has typically
$8\cdot10^4$ atoms, a critical temperature of $0.5~\mu{\rm K}$ and
is at $300~\mu{\rm m}$ from the chip surface. The trap has a
longitudinal frequency of $46~{\rm Hz}$, the radial trapping
frequency is $950~{\rm Hz}$. Substantial heating of the atom chip
during the chip loading phase and the need for pumping out all the
polluting gas emitted by the dispenser impose a $15~{\rm s}$
recovery time. Thus a complete experimental cycle has a duration
of $23~{\rm s}$.

All subsequent manipulations are made $0.7~\rm{ms}$ after turning
off the magnetic trap to guarantee bias field homogeneity (at
$3.1~{\rm Gauss}$, corresponding to an energy splitting of the
Zeeman states of about $2.2~{\rm MHz}$) and strongly reduce the
effects of atomic collisions. To drive the Raman transition, we
derived from a diode laser source two coherent beams, one of them
passing an electro-optic modulator (EOM) plus Fabry-P\'{e}rot cavity
combination to shift its frequency by $6.834~{\rm GHz}$. The laser
beams with mutually perpendicular linear polarizations are red
detuned by about $18~\rm GHz$ and $25~{\rm GHz}$ from the nearest
resonance of the D2 line at $780~{\rm nm}$. The beams have their
waist of about $70~\mu{\rm m}$ on the atomic sample and are
propagating collinearly in the direction parallel to the applied
bias field. Since the Land\'{e} factors of the two hyperfine states
have opposite signs (cf.\ Fig.~\ref{fig:levels} of the main text),
in the presence of the bias magnetic field the two laser beams can
be resonant only with a single pair of levels at a time. We choose
the frequency difference such that the connected levels are
$|2,0\rangle$ and $|1,0\rangle$. To maximize the transfer efficiency
the beam intensities are chosen to be equal. We can induce
oscillations with frequencies up to $400~\kHz$. Off resonant
scattering from the excited states damps these oscillations in
around $50~\mus$. The illumination with Dissipative Light can remove
all atoms from the $F=1$ state in $0.6~\mus$. Note that the emission
of a spontaneous photon is enough to kick the atoms out of the BEC.
Therefore this interaction is, for our purpose, completely
destructive.

To record the number of atoms in each of the $m_F$ states of the
$F=2$ hyperfine state we apply a Stern-Gerlach method. After $1~{\rm
ms}$ of free expansion in addition to the homogeneous bias field an
inhomogeneous magnetic field is applied along the quantization axis
for $10~{\rm ms}$. This causes the different $m_F$ states to
spatially separate. After a time of $23~{\rm ms}$ of expansion a
standard absorption imaging sequence is executed (see
Fig.~\ref{fig:si1}, lower part). By performing interleaved control
measurements (by not turning on the RF field) we distinguish the
fraction of atoms lost from the $F=2$ manifold due to leakage out of
the protected two-level system from atoms lost due to residual
scattering of the Raman and Dissipative Light beams. Losses due to
collisional heating induced by the Dissipative Light are up to about
$50\%$ during the experimental sequence. The data is corrected not
to include losses due to scattering.

{\bf Density matrix two-level model.}
The two-level model expressed in terms of the density matrix $\rho$
is
\begin{equation}
    \begin{array}{rcl}
        \dot{\rho}_{\alpha\alpha} & = & i \, \Omega\, (
        \rho_{\alpha\beta} - \rho_{\alpha\beta}^* ) \\

        \dot{\rho}_{\beta\beta} & = & -i \, \Omega\, (
        \rho_{\alpha\beta} - \rho_{\alpha\beta}^* )
        - 2\, \Gamma_{\rm loss}\, \rho_{\beta\beta} \\

        \dot{\rho}_{\alpha\beta} & = & i \, \Omega\, (
        \rho_{\alpha\alpha} - \rho_{\beta\beta} )
        - (\Gamma_{\rm loss} + 2\, \gamma_{\rm deph})\,
        \rho_{\alpha\beta}\, ,
    \end{array}
    \label{eqn:density}
\end{equation}
where $\rho_{ii}$, ($i=\alpha, \beta$) is the population of the
levels $m_F = 2$ and $m_F = 1$, respectively, and $\rho_{\alpha
\beta}$ is their coherence. The error intervals of the fit
are taken as the range where the squared sum of the fit residuals
worsens by $10\,\%$ under variation of one parameter only.

{\bf QZD via four different protocols.}
To induce QZD four different protocols were exploited. Here we
report details on the experimental parameters for each protocol.

In the first protocol, {\it discrete measurements}, a sequence of
double-pulses is applied to the atoms. We first apply a $0.8~\mus$
$\pi$-pulse (i.e.\ a pulse that transfers all the atoms that might
be present in the $|2,0\rangle$ state to the $|1,0\rangle$ state)
with the Raman beams. Immediately after this pulse, we apply for
$0.6~\mus$ the Dissipative Light. As a result of this sequence, any
atom that would have been initially present in $|2,0\rangle$ would
be detected. We repeat this sequence every $2.2~\mus$.

For the second protocol, {\it continuous measurements}, we apply
both the Raman beams and the Dissipative Light continuously. The
intensities are chosen such as to obtain a Raman induced coupling of
$2\pi\,250~\kHz$ and a dissipative loss rate of $2\pi\,450~\kHz$.
The intensity of the Dissipative Light was chosen such as to induce
a loss rate only slightly above the Raman coupling rate. This was
done in order to avoid blocking the Raman transition by a
Dissipative Light induced QZE, while still implementing an effective
measurement scheme.

Our third protocol, {\it unitary kicks}, consists of sending a
sequence of isolated pulses onto the atoms. The Raman beams
$\pi$-pulses have a duration of $0.7~\mus$ and are repeated every
$2.2~\mus$.

The fourth protocol, {\it continuous unitary coupling}, relies on
the sole usage of the Raman beams. They are tuned to a coupling
strength of $2\pi\,136~\kHz$.

With the chosen parameters the lifetimes $2\pi / \Gamma_{\rm loss}$
are observed to be in the range $0.7~\mus$ to $1.0~\mus$. The
decoherence rates $\gamma_{\rm deph}$ are found to lie between
$2\pi\, 0.3~\kHz$ and $2\pi\, 0.8~\kHz$.


\section{Acknowledgments}
This work was supported by the Seventh Framework Programme for
Research of the European Comission, under FET-Open grant MALICIA
(265522), and QIBEC (284584). The work of F.C. has been supported by
EU FP7 Marie–Curie Programme (Career Integration Grant) and by
MIUR-FIRB grant (RBFR10M3SB). We thank M. Inguscio for fruitful
discussions and continuous support. We wish to thank M.\
Schramb\"{o}ck (Atominstitut, TU-Wien) at the ZMNS (TU-Wien) for the
realization of the atom chip. QSTAR is the MPQ, LENS, IIT, UniFi
Joint Center for Quantum Science and Technology in Arcetri.

\section{Author Contributions}
F.S., I.H., S.C., C.L. and F.S.C. carried out the experiment; F.C.
and A.S. the theoretical work. F.S. and F.C. performed the numerical
simulations. F.S.C. and A.S. supervised the project. All authors
contributed to the discussion and analysis of the results and the
writing of the manuscript.

\end{document}